\documentclass[journal=jacsat,manuscript=article]{achemso}

\usepackage{chemformula} 
\usepackage[T1]{fontenc} 
\usepackage{dcolumn}
\usepackage{graphicx,amssymb,textcase,setspace,fancyhdr,enumerate}
\usepackage[utf8]{inputenc}
\usepackage{subcaption}
\usepackage{caption}
\usepackage{float}
\usepackage{url}
\usepackage{hyperref}
\usepackage{array}
\usepackage{booktabs}
\usepackage{listings}
\usepackage{lmodern}
\usepackage{mathpazo}
\usepackage{microtype}
\usepackage{geometry}
\usepackage{xkeyval}
\SectionNumbersOn

\newcommand{\angstrom}{\mbox{\normalfont\AA}}

\author{Márton Guba }
\affiliation[Budapest University of Technology and Economics]
{Department of Inorganic and Analytical Chemistry, Faculty of Chemical Technology and Biotechnology,  Budapest University of Technology and Economics, Műegyetem rkp. 3., H-1111 Budapest, Hungary\\}
\alsoaffiliation[Budapest University of Technology and Economics]
{HUN-REN-BME Computation driven chemistry research group, Budapest University of Technology and Economics, Műegyetem rkp. 3., H-1111 Budapest, Hungary\\}
\author{Tibor Höltzl}
\affiliation[Budapest University of Technology and Economics]
{Department of Inorganic and Analytical Chemistry, Faculty of Chemical Technology and Biotechnology,  Budapest University of Technology and Economics, Műegyetem rkp. 3., H-1111 Budapest, Hungary\\}
\alsoaffiliation[Budapest University of Technology and Economics]
{HUN-REN-BME Computation driven chemistry research group, Budapest University of Technology and Economics, Műegyetem rkp. 3., H-1111 Budapest, Hungary\\}
\alsoaffiliation[Furukawa Electric Institute of Technology]
{Furukawa Electric Institute of Technology Ltd., Nanomaterials Science Group, Budapest, Hungary}
\email{tibor.holtzl@furukawaelectric.com}
\title{Electrode Potential Dependent Differential Capacitance in Electrocatalysis: a Novel, Ab Initio Computational Approach}

\keywords{American Chemical Society, \LaTeX}

\begin{document}

\begin{tocentry}

\begin{figure} [H]
    \centering
    \includegraphics[scale=0.21]{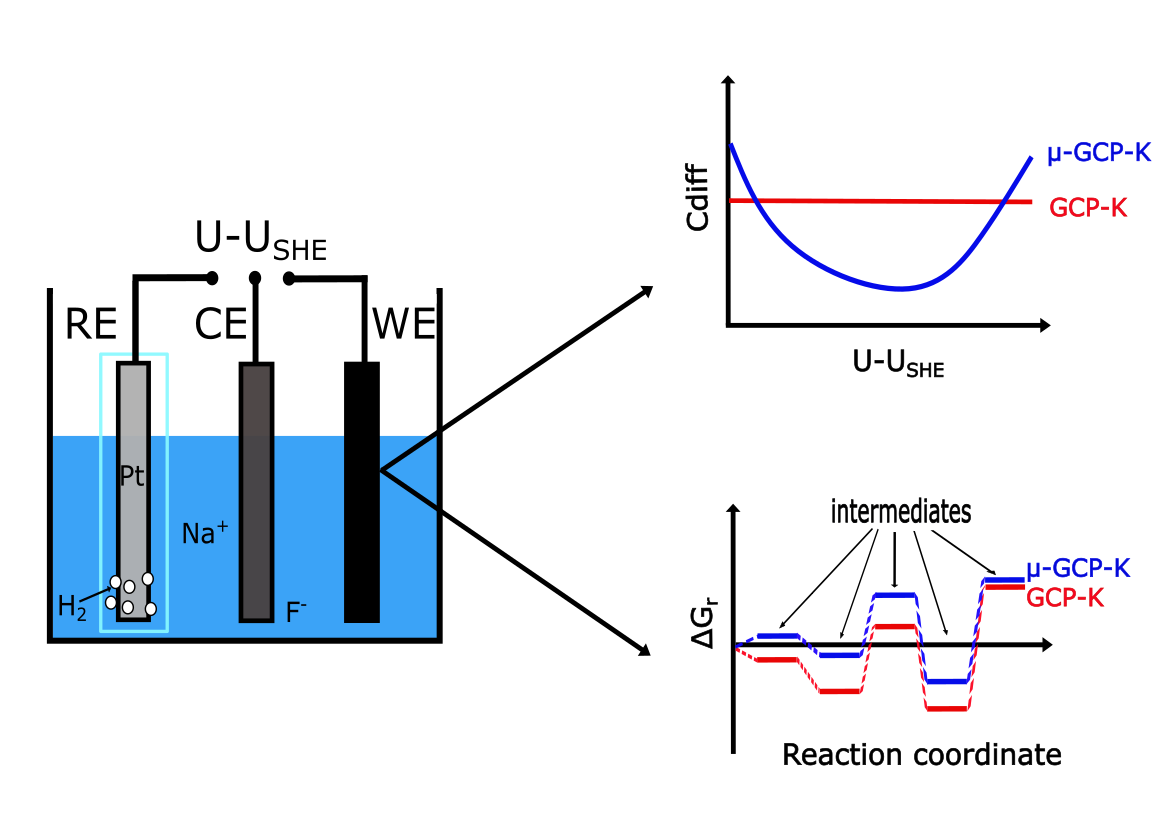}
\end{figure}

\end{tocentry}
\newpage
\begin{abstract}
    As interest in nanomaterials grows, ab initio simulations play a crucial role in designing electrochemical catalysts. Electrochemical reactions depend on electrode potential, highlighting the importance of the grand canonical representation, especially when integrated with Density Functional Theory.  The Grand Canonical Potential - Kinetics (GCP-K) method is a valuable approach for determining electrocatalytic reaction mechanisms and kinetics rooted in quantum mechanics, relying on assumptions of quadratic free energy dependence on charge and a constant differential capacitance-potential relationship.  However, it is known that differential capacitance is potential-dependent in several practical electrocatalysts. Here we present $\mu$-GCP-K, a practical approach which makes no assumptions about the relationships between thermodynamic and electrochemical properties. We demonstrate the method's efficiency by computing the surface charge density and differential capacitance of graphene, further emphasizing the importance of accurately calculating the thermodynamic stability of reaction intermediates in $CO_2$ electroreduction, while also showing the role of potential-dependent differential capacitance.
\end{abstract}
\begin{figure} [H]
      \centering
	\begin{subfigure}{0.48\linewidth}
		\includegraphics[width=\linewidth]{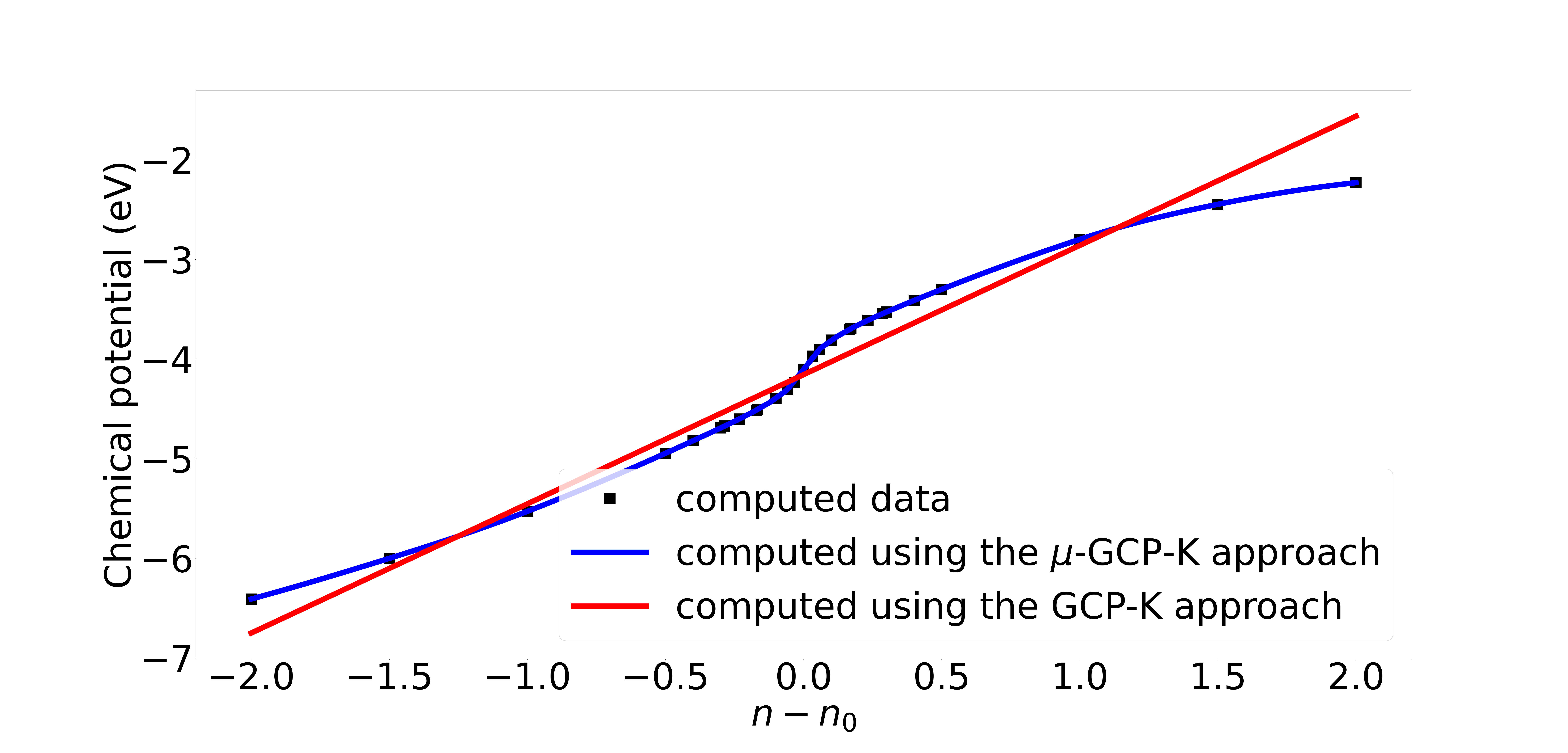}
		\caption{The $\mu \left (\Delta n\right)$ characteristic of graphene.}
		\label{fig_muF_graphene_1}
	   \end{subfigure}
	   \begin{subfigure}{0.48\linewidth}
		\includegraphics[width=\linewidth]{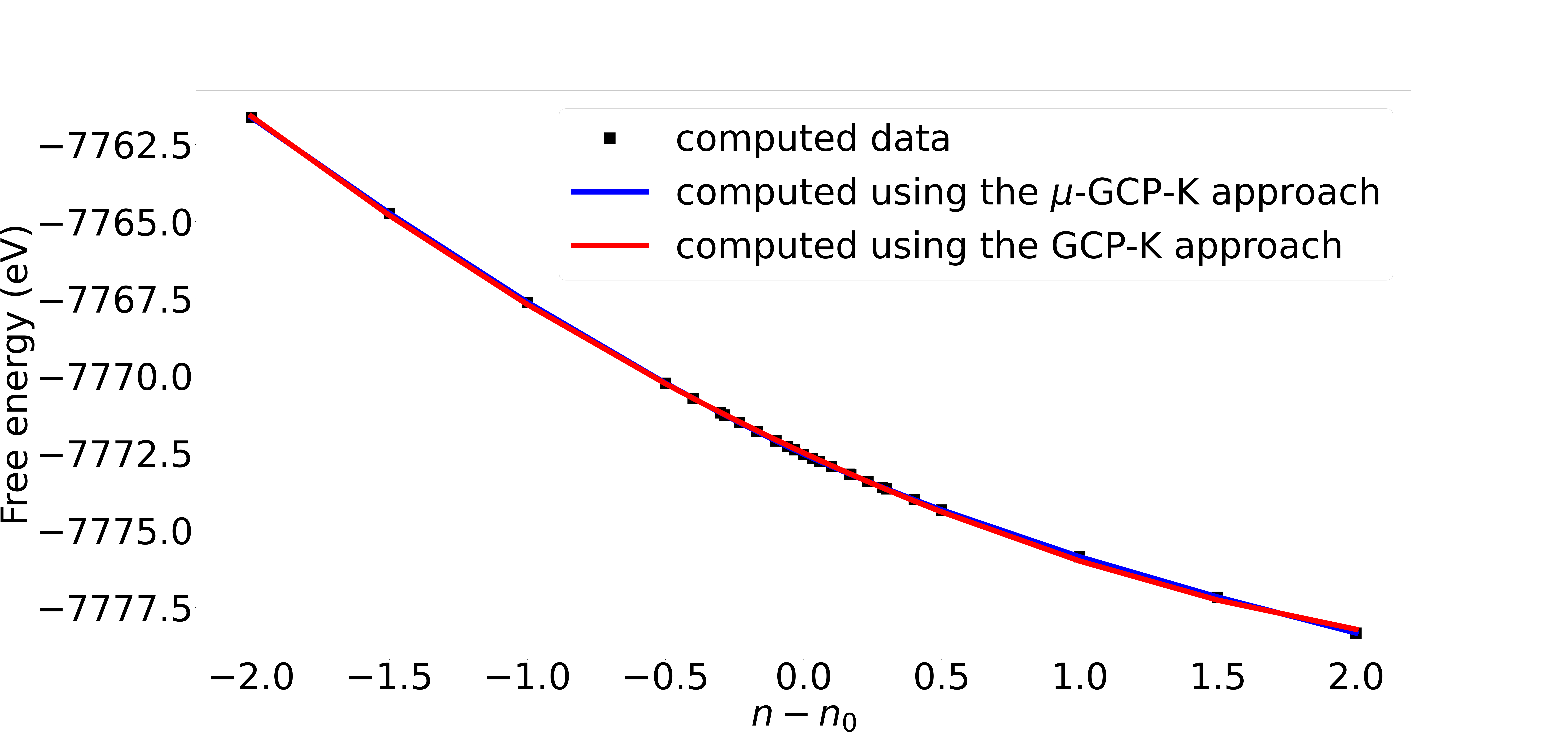}
		\caption{The $F \left (\Delta n\right)$ characteristic of graphene.}
		\label{fig_muF_graphene_2}
	    \end{subfigure}
	\caption{The electrochemical potential and free energy of graphene in terms of $\Delta n$. Fittings were carried out on the computed data proposing quadratic (red) and general (blue) $F \left (\Delta n\right)$ relation.}
	\label{fig_muF_graphene}
\end{figure}
\begin{figure} [H]
      \centering
	\begin{subfigure}{0.45\linewidth}
		\includegraphics[width=\linewidth]{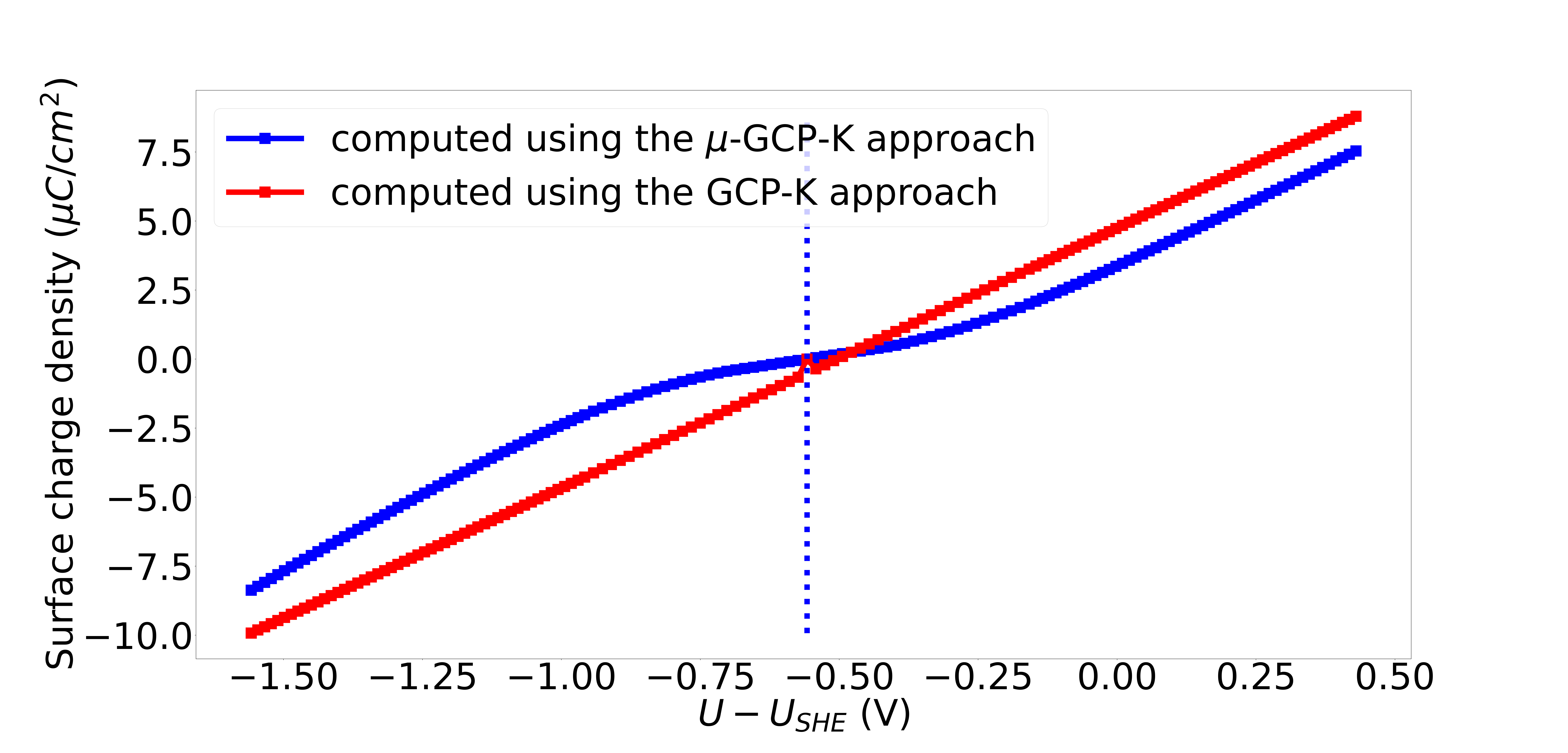}
		\caption{The $\sigma\left(U-U_{SHE} \right)$ characteristic of graphene.}
		\label{fig_nC_graphene_1}
	   \end{subfigure}
	   \begin{subfigure}{0.45\linewidth}
		\includegraphics[width=\linewidth]{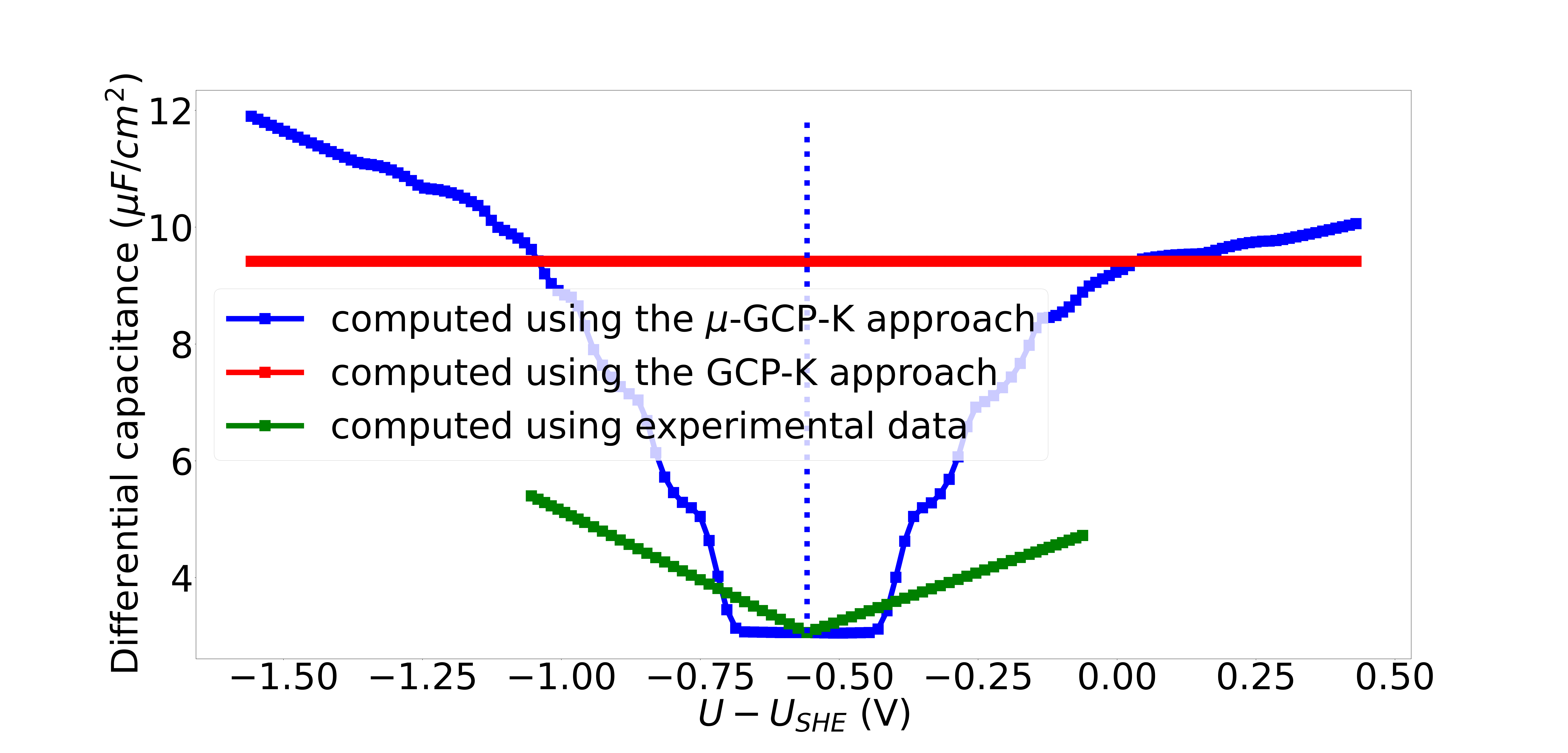}
		\caption{The $-C_{diff} \left(U-U_{SHE} \right)$ characteristic of graphene.}
		\label{fig_nC_graphene_2}
	    \end{subfigure}
	\caption{The surface charge density and the differential capacitance in terms of the electrode potential. The values computed using quadratic assumption of free energy are denoted by red, the ones evaluated by the $\mu$-GCP-K method are presented with blue and the ones reconstructed based on the experimental data of Zhang et al. \cite{graphene_Cdiff_ref_31} are denoted by green rectangles. Blue, vertical dashed line shows the PZC of the model, calculated by the $\mu$-GCP-K approach.}
	\label{fig_nC_graphene}
\end{figure}

\newpage
Ab initio simulations have become essential in the design of electrocatalysts \cite{sym_catal_48, sym_catal_49, sym_catal_50} by modelling and predicting their reactivities under electrochemical conditions. The electrode-electrolyte interface \cite{electrode_eletrolyte_45, electrode_eletrolyte_46, electrode_eletrolyte_47} plays a significant role in these processes. Here, charge accumulation at the interface can be modelled as an Electrochemical Double Layer (EDL) \cite{EDL_63, EDL_64} leading to capacitive behaviour. Since electrochemical reactions occur through charge transfer, the capacitance of the EDL, determined by the combined effects of the electrode and electrolyte, is a crucial property that influences the energetics of these reactions \cite{Cdiff_reactivity_78,EDL_reac_65, EDL_reac_66}. This capacitance cannot be assumed to be constant in terms of the electrode voltage, particularly when quantum effects dominate near the Fermi level \cite{EDL_64, Binninger_59}. Specifically, we investigate how the electrode potential affects the differential capacitance and alters energetic relationships by adjusting the charging of the electrode. \\ 
One of the most popular methods for simulation of the effect of electrode potential is the Computational Hydrogen Electrode (CHE) model, developed by Nørskov et al. \cite{CHE_15}, which describes the evolution of the thermodynamic stability of the reaction intermediates without considering the electrode potential dependent charging of the electrode. On the other hand, the Grand-Canonical Density Functional Theory (GC-DFT) \cite{GCDFT_51, GCDFT_52, GCDFT_53} addresses this by modifying the occupations of the electronic bands in the electrode, filling or depleting them up to energies determined by the electrode potential relative to the Fermi level of the neutral system. The problem is defined in the framework of the grand-canonical ensemble, i.e. the number of electrons, $n$ can vary according to the electrode potential $U$ (proportional to the target chemical potential). In fact, Jinnouchi shows, that methods discussed so far are all based on equations defined in the grand-canonical ensemble with different levels of approximation \cite{GCNDFT_84}. Although GC-DFT computations are self-consistent and account for electronic relaxations, the target chemical potential for filling the electrons is calculated in the neutral case, without considering the changes in band dispersion due to orbital relaxation in the charged electrode. The Grand Canonical Potential-Kinetics (GCP-K) method of Goddard et al. \cite{GCPK_18, GCPK_19, GCPK_20} goes one step further, as it treats the effects of the charged system self-consistently. Variations in $n$ affect both the electronic and thermodynamic properties of the electrode, indicating that not only does its reaction free energy, $G_{r}$, depend on the electrode potential, but so does its grand canonical potential, $G$. This is directly reflected by the following three terms establishing the formalism of the GCP- K. 
\begin{equation}
            G(n,U)=F(n)-ne \left (U_{SHE}-U \right).
            \label{GCPK_as_1}
        \end{equation}
        Here,  $F$ is the thermodynamic free energy  (referred to as free energy hereafter), $n$ is the number of electrons, $e$ is the unit charge, while the potential $U$ is measured from the reference potential of the standard hydrogen electrode (SHE), $U_{SHE}=-4.66$ V \cite{USHE_14}, determined by correlating measured Potential of Zero Charge (PZC) values of metallic electrodes with ab initio computed electronic chemical potential values\cite{CANDLE_60}. 
        A fixed electrode potential leads to the variation of $n$ until a constant charge state is obtained, thus the equilibrium condition can be expressed as follows.
\begin{equation}
            G(U)= \underset{n}{min} \left \{ F(n) -ne \left (U_{SHE}-U \right) \right \}.
            \label{GCPK_as_2}
        \end{equation}   
         Several studies indicate \cite{quadU_16, quadU_17}, that $G$ should be at least quadratic in $U$ due to capacitive coupling between the electrode and the electrolyte. Hence, GCP-K fits the electron-number dependent free energy as
\begin{equation}
        F(n)=a \left(n-n_{0} \right)^{2}+b \left(n-n_{0} \right) +c,
        \label{GCPK_as_3}
    \end{equation} 
    where $a$, $b$ and $c$ are coefficients and $n_{0}$ represents the number of electrons in neutral case.  While Equations (\ref{GCPK_as_1}) and (\ref{GCPK_as_2}) provide general insights, Equation (\ref{GCPK_as_3}) limits the dependence of free energy on the number of electrons to at most second order, implying a potential independent differential capacitance. However, this assumption does not hold in several practical cases.The GCP-K calculations start with the quadratic function fitting to the ab initio computed free energy values, using the form presented in Equation (\ref{GCPK_as_3}). Without going into the details \cite{GCPK_18, GCPK_19, GCPK_20}, the back-substitution into Equation (\ref{GCPK_as_1}) and the evaluation of Equation (\ref{GCPK_as_2}) result $G(U)$ as:
\begin{equation}
    G(U)=\frac{C_{diff}}{2} \left(U-U_{PZC}\right)^{2}+n_{0}eU+F_{0}-n_{0}\mu_{SHE}.
    \label{G_U}
\end{equation}
Here, $C_{diff}$ is the differential capacitance of the system, $U_{PZC}$ is the PZC, where $n(U_{PZC})=n_{0}$, $F_{0}$ is the free energy of the system at $U=U_{PZC}$ and $\mu_{SHE}=eU_{SHE}$ is the electrochemical potential of the SHE. We emphasize  that $C_{diff}$ contains contributions corresponding both to the electrode and the electrolyte which is represented in the linear, implicit solvation methodology, specifically the charge-asymmetric nonlocally-determined local-electric (CANDLE) model \cite{CANDLE_60},  used to simulate the electrode-electrolyte interaction. In this context, the effect of dilute, dissolved ions is described by the Poisson-Boltzmann equation \cite{USHE_14}. $C_{diff}$, $U_{PZC}$ and $F_{0}$ are related to the coefficients defined in Equation (\ref{GCPK_as_3}):
\begin{equation}
    C_{diff}=e\frac{\partial{n}}{\partial{U}}=-\frac{e^{2}}{2a}, U_{PZC}=\frac{1}{e} \left(\mu_{SHE}-b\right), c=F_{0}
    \label{coeffs}
\end{equation}
The number of electrons $n(U)$ and the chemical potential $\mu (n)$ are computed as:
\begin{equation}
    n=n_{0}+\Delta n = n_{0} + \frac{C_{diff}}{e} \left (U-U_{PZC} \right),
    \label{n}
\end{equation}
\begin{equation}
    \mu(n)=\frac{\partial{F}}{\partial{n}}=2a\left(n-n_{0} \right) + b.
    \label{mu}
\end{equation}
 Equation (\ref{mu}) suggests that the electrochemical potential of the system should vary linearly in terms of $\Delta n$, while Equation (\ref{GCPK_as_3}) indicates a quadratic dependence of free energy on the electron number.\\ 
This approach is effective for simulating ideal metallic electrodes, however, it may not yield the same accuracy for systems with more complex electronic structures, such as nanoparticles \cite{nanopart_54}, clusters \cite{clus_55} or semi-metals \cite{semimet_56}. Most notably, the quadratic $F(n)$ assumption of GCP-K leads to a potential-independent differential capacitance, while the importance of non-constant $C_{diff}$ has been shown to play important role in different electrochemical processes \cite{eff_diff_cap_57, eff_diff_cap_58, Binninger_59}. \\
One of the notable such case is graphene, which due to its high electric conductivity \cite{graphene_cond_21, graphene_cond_22}, facile synthesis \cite{graphene_synth_23, graphene_synth_24} and mechanical elasticity \cite{graphene_prop_25}, serves as promising ingredient in electrocatalysts and is also often applied as model system in theoretical studies. However, graphene is known for its potential-dependent differential capacitance \cite{graphene_Cdiff_ref_31, graphene_Cdiff_ref_32, graphene_Cdiff_ref_33, graphene_Cdiff_ref_34}, while GCP-K yields a constant value. \\
Here, we extend the GCP-K method to accommodate cases with non-constant differential capacitance and non-quadratic free energy in terms of the electrode potential. By analyzing $\mu(n)$ derived from DFT computations, we developed a more general approach (referred to as $\mu$-GCP-K), based solely on Equations (\ref{GCPK_as_1}) and (\ref{GCPK_as_2}). To express the equilibrium form of Equation (\ref{GCPK_as_1}), the variation of $n$ at a given $U$, $\Delta n^{*}$ should be determined:
  \begin{equation}
     G(U)=F\left(n_{0}+\Delta n^{*}\right)-e\left(n_{0}+\Delta n^{*}\right)\left (U_{SHE}-U \right).
     \label{GU}
 \end{equation}
$\Delta n^{*}$ can be computed by solving the variational problem, presented in Equation (\ref{GCPK_as_2}): 
 \begin{equation}
    \frac{\partial{G}}{\partial{n}}=0 \leftrightarrow \frac{\partial{F}}{\partial{n}}-e \left (U_{SHE}-U \right) = 0 \leftrightarrow \mu(\Delta n^{*})-e \left (U_{SHE}-U \right)=0.
     \label{delta_n_*}
 \end{equation}
The root finding problem shown by Equation (\ref{delta_n_*}), is solved numerically in our method by interpolating the computed $\mu(n)$ values. We compute $\mu(n)$ for various values of $\Delta n$ on an $\Delta n \in \left [-2.0, 2.0 \right ]$ grid with $0.5$ resolution ($0.1$ for the graphene close to the Fermi level) and apply piecewise cubic interpolation to the results. The same grid is applied to compute $F(n)$ and using $\Delta n^{*}$, determined based on Equation (\ref{delta_n_*}), $G(U)$ is evaluated. The rearrangement of Equation (\ref{delta_n_*}) with the $\mu(\Delta n=0)$ term provides the PZC, while the potential dependent differential capacitance can be computed according Equation (\ref{coeffs}) in terms of $U$. Both the GCP-K and $\mu$-GCP-K begin with DFT computations carried out for neutral and charged systems. The key distinction is that the calculations of the GCP-K do not account for how different electronic structures influence the differential capacitance and grand canonical potential, nor do they analyze the relationship between $F(n)$ and $\mu(n)$ described by Equation (\ref{mu}). In contrast, $\mu$-GCP-K starts with the investigation of $\mu(n)$, which more directly reflects the electronic properties of the model compared to $F(n)$. This focus allows for a more fundamental simulation of the system's properties under electrochemical conditions.\\
We first evaluate the $\mu$-GCP-K approach for its accuracy in determining the differential capacitance of graphene. DFT computations with the jDFTx software, developed by Sundararaman et al. \cite{jDFTx_85}, were performed on a $5$x$5$ graphene supercell model with the PBE functional \cite{PBE_70}, D3 dispersion correction of Grimme\cite{D3_71}, plane wave (PW) basis set with $20$ a.u. and charge density with $100$ a.u. cutoff energies (default values) and Monkhorst-Pack momentum space sampling scheme \cite{Monkhorst-Pack_72} (k-point set) with size $(5,5,1)$. The chemical potential and the free energy of graphene are depicted in Figure \ref{fig_muF_graphene}. 
Considering the $\mu \left (\Delta n \right)$ relation, the fittings based on linear and general $\mu \left (\Delta n\right)$ functions show major differences. This can be numerically expressed in terms of the Root Mean Square (RMS) of the deviance between the computed and fitted data, which is $RMS=0.23$ eV corresponding to the data derived from the linear assumption and technically zero (in the order of $10^{-16}$ eV due to numerical fitting error), when the $\mu$-GCP-K method is applied. The situation improves for the free energy, where $RMS=0.06$ eV was obtained assuming quadratic relation of GCP-K for $F\left ( \Delta n \right)$  (still $RMS \approx 0$ eV with the $\mu$-GCP-K). We computed the surface charge density, $\sigma=\frac{-n+n_{0}}{A}$, with surface area, $A=131.148$ $\angstrom^{2}$ to illustrate the consequences of the linear charge-dependence of $\mu$. For the sake of better comparison with previously reported experimental values, the differential capacitance, $-C_{diff}$ in terms of the electrode potential $U-U_{SHE}$ was also determined and presented in Figure \ref{fig_nC_graphene}. 
Several previous works discuss the theoretical and experimental studies on the dependence of the charge and the differential capacitance on the electrode potential of graphene-based electrodes \cite{graphene_Cdiff_ref_31, graphene_Cdiff_ref_32, graphene_Cdiff_ref_33, graphene_Cdiff_ref_34, graphene_Cdiff_73, graphene_Cdiff_74, graphene_Cdiff_75, graphene_Ndopedgraphene_Cdiff_80}. These confirm the non-linear dependence of the charge density and the non-constant characteristic of the differential capacitance on the electrode voltage. Zhan et al.\cite{graphene_Cdiff_ref_79}, Zhang et al. \cite{graphene_Cdiff_ref_31}, Ochoa-Calle et al. \cite{graphene_Ndopedgraphene_Cdiff_80} and Kasamatsu et al. \cite{graphene_Cdiff_ref_32} computed the surface charge density of a graphene electrode under electrochemical conditions and derived a similar evolution as a function of the potential, consistent with the calculations obtained using the $\mu$-GCP-K approach, as shown in Figure \ref{fig_nC_graphene_1}. These confirm that the linear $\sigma(U)$ relation does not reflect the realistic electrochemical properties of graphene. Graphene, characterized by the presence of the Dirac-cone in its band structure, is known to exhibit a V-shaped Density of States (DOS) in energy space near the Fermi level in the neutral case \cite{graphene_prop_25, graphene_prop_37}. The DOS of graphene can be related to its $C_{diff}$ shown and used by several studies \cite{graphene_Cdiff_ref_33,graphene_Cdiff_73, graphene_Cdiff_74, graphene_Cdiff_75} resulting the dominance of its contribution to the $C_{diff}$ characteristics close to the PZC \cite{graphene_DOS_Cdiff_83,graphene_Cdiff_ref_32, graphene_Cdiff_ref_34}. This was also confirmed by the measurements of Zhang et al. \cite{graphene_Cdiff_ref_31}, Ji et al. \cite{graphene_Cdiff_ref_33}, Wang et al. \cite{graphene_Cdiff_exp_76} and Xing et al. \cite{graphene_Cdiff_exp_77}. The results presented in the work of Zhang et al. \cite{graphene_Cdiff_ref_31} (linear fit with slopes $3.35$ and $-4.72$ $\frac{\mu F}{cm^{2}V}$ for $U>PZC$ and $U<PZC$, respectively) were used to reconstruct these experimental data and clearly show, that the computed results of the $\mu$-GCP-K theory cover more correctly the recorded $C_{diff}(U)$ curve close to the PZC compared to that of the quadratic assumption.
\begin{table} [H]
\centering
\caption{The differential capacitance of graphene measured and computed at $ U-U_{SHE}=PZC$.}
\label{tab_Cdiff_graphene} 
\begin{tabular}{cc}
\hline
Source & $C_{diff}$ at $U-U_{SHE}=PZC$  $\left(\frac{\mu F}{cm^{2}}\right)$\\
\hline
$\mu$-GCP-K          & 3.5  \\                                                                     
quadratic fitting of $F(n)$   & 9.5  \\                                                        
computational work of Zhan et al. \cite{graphene_Cdiff_ref_79} & 1.5 \\                            
experimental work of Zhang et al. \cite{graphene_Cdiff_ref_31} & 1.5  \\                                    
experimental work of Ji et al. \cite{graphene_Cdiff_ref_33}    & 2.5-3.5 \\                                 
experimental work of Wang et al. \cite{graphene_Cdiff_exp_76}   & 0.8   \\                                 
experimental work of Xing et al. \cite{graphene_Cdiff_exp_77}    & 6.7   \\                                 
\hline
\end{tabular}
\end{table}
As Table \ref{tab_Cdiff_graphene} illustrates, the computed value of $C_{diff}\left ( U-U_{SHE}=PZC \right)$ using the $\mu$-GCP-K approach is usually closer to the measured ones than that of the one assuming quadratic $F(n)$ relation. Some differences may occur originated from the various electrolyte types and concentrations. Similarly to Xing et al., we applied  $1$ mol/L NaF electrolyte with $H_{2}O$ solvent in our simulations, Ji et al. used a $6$ mol/L KOH \cite{graphene_Cdiff_ref_33}, while Zhang et al. employed a $2$ mol/L NaCl aqueous solution \cite{graphene_Cdiff_ref_31} (the capacitance measurements of Wang et al. were not performed with electrochemical setup). We performed additional simulations with NaCl solvate (KOH is not accessible in the program) for the sake of comparison and found minor deviation ($\Delta C_{diff} \approx 0.04$ $\frac{\mu F}{cm^ {2}}$) compared to the NaF case. The PZC is also an important electrochemical property of the electrode providing the sign of the net charge at a given potential. Similarly to the $C_{diff}$, we listed computed and measured values in Table \ref{tab_PZC_graphene}.
\begin{table}[H]
\centering
\caption{The absolute measured and computed PZC of graphene.}
\label{tab_PZC_graphene}
\begin{tabular}{cc}
\hline
Source               & Absolute PZC (V) \\ 
\hline
$\mu$-GCP-K          & 0.56                                              \\ 
quadratic fitting of $F(n)$    & 0.50                                              \\ 
experimental work of Zhang et al. \cite{graphene_Cdiff_ref_31} & 0.32                    \\ 
\hline
\end{tabular}
\end{table}
Table \ref{tab_PZC_graphene} indicates $\sim 0.2$ V difference between the measured PZC of Zhang et al. and the computed values, which can be originated from several sources, e.g. the generalized gradient approximation \cite{PBE_70} employed during the computation \cite{GGA_acc_86}, furthermore the deviance strongly depends on the reference value of the SHE which is used to be in the range of $\left [4.4, 4.9 \right ]$ V \cite{SHE_35}. This means at least $0.5$ V uncertainty, which makes difficult to compare the computed and measured values.\\       
We also investigate the chemical implications of the electrochemical potential-dependent differential capacitance, as described by the $\mu$-GCP-K method in the copper-doped graphitic carbon nitride ($g-C_{3}N_{4}$-$Cu$) model system, using the same computational method as discussed for graphene. Similar materials are subject of intensive research as single-atom catalysts for carbon dioxide electrocatalytic reduction ($CO_2RR$) \cite{GCN_Cu_26, GCN_Cu_27, GCN_Cu_28}. Graphitic carbon nitride ($g-C_{3}N_{4}$) as nitrogen-rich formal derivative of graphene, is an attractive candidate to support metal atoms and particles due to its high surface area and several possible binding sites. It is known to be a semiconductor with an approximate band gap of $2.7$ eV \cite{GCN_gap_61, GCN_gap_62}, which results in a discontinuity in the occupation of electronic bands, as illustrated in Figure \ref{fig_g-C_{3}N_{4}_mu}.
\begin{figure} [H]
      \centering
	\begin{subfigure}{0.48\linewidth}
		\includegraphics[width=\linewidth]{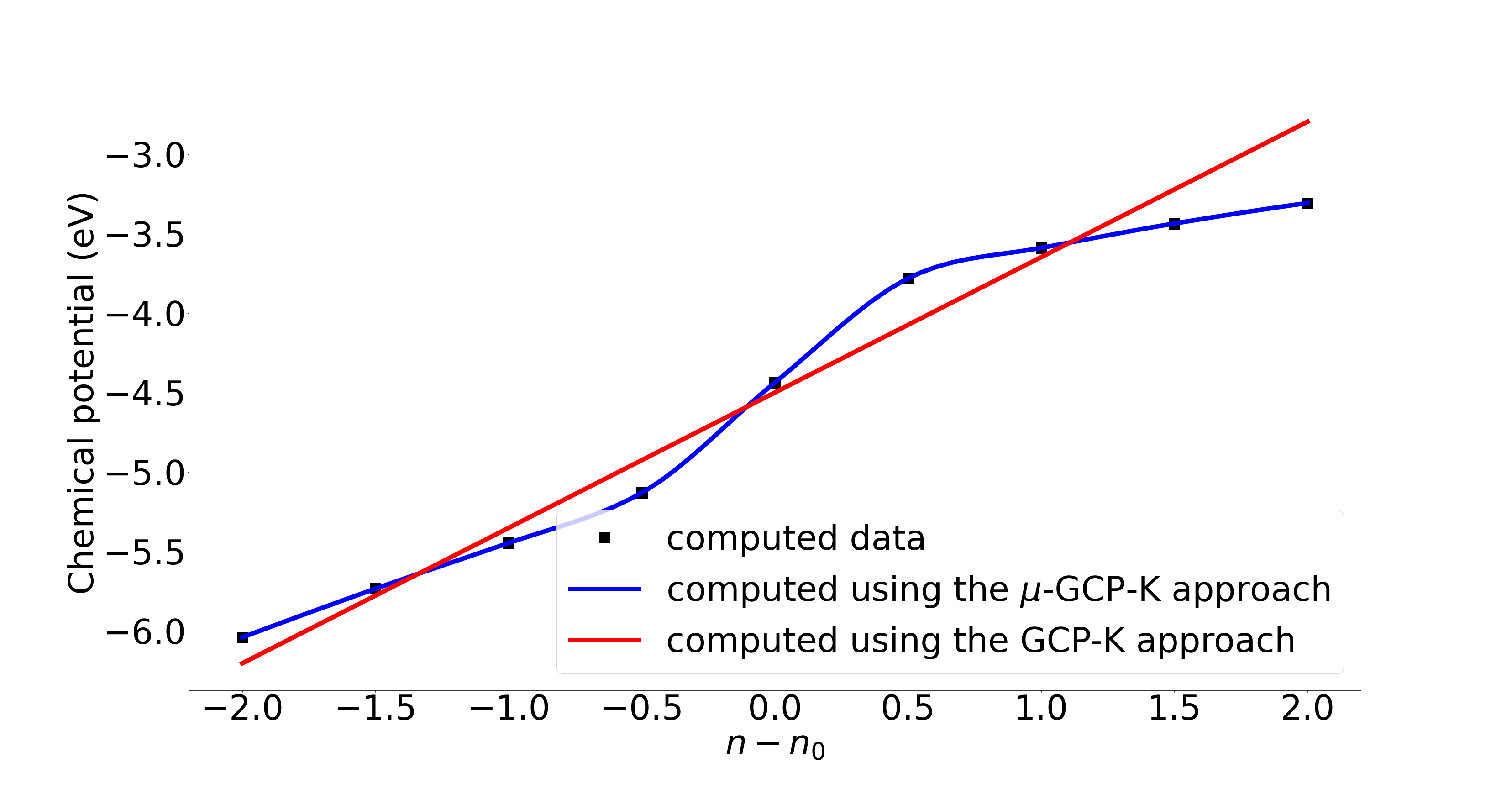}
		\caption{The $\mu \left (\Delta n\right)$ characteristic of $g-C_{3}N_{4}$.}
		\label{fig_g-C_{3}N_{4}_mu}
	   \end{subfigure}
	   \begin{subfigure}{0.488\linewidth}
		\includegraphics[width=\linewidth]{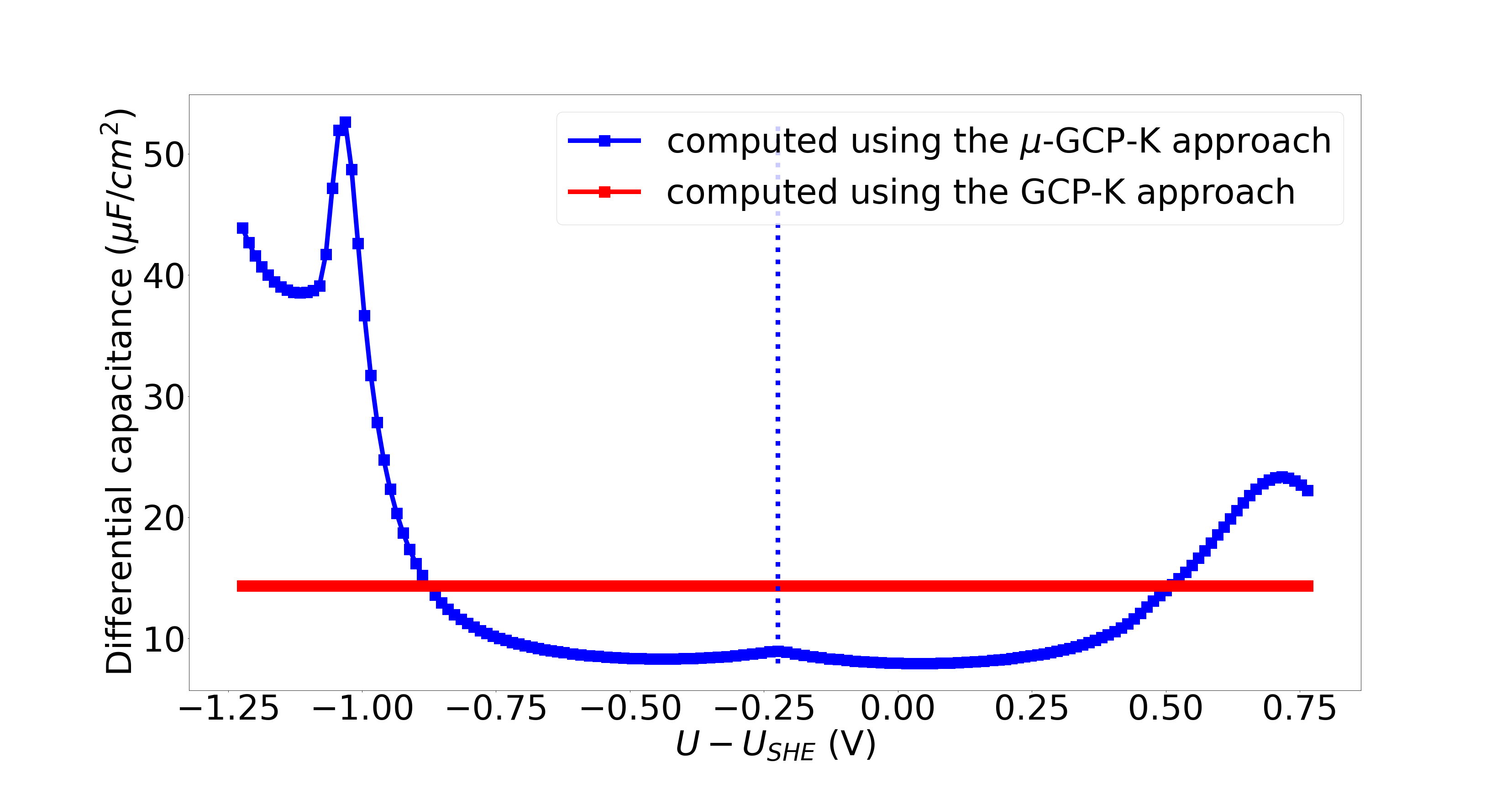}
		\caption{The $-C_{diff} \left(U-U_{SHE} \right)$ characteristic of $g-C_{3}N_{4}$.}
		\label{fig_g-C_{3}N_{4}_Cdiff}
	    \end{subfigure}
	\caption{The electrochemical potential and differential capacitance of $g-C_{3}N_{4}$. Blue, vertical dashed line shows the PZC of the model in the $-C_{diff} \left(U-U_{SHE} \right)$ characteristic, calculated by the $\mu$-GCP-K approach.}
	\label{fig_g-C_{3}N_{4}_mu_Cdiff}
\end{figure}
The increase at $\Delta n \approx 0$ in the $\mu \left(\Delta \right)$ relation is a consequence of the band gap, resulting in a root mean square deviation (RMSD) of $0.23$ eV from that using the linear $\mu \left(\Delta \right)$ approximation. Unlike what has been observed for graphene, the slope of the $\mu \left(\Delta \right)$ curve is asymmetric around the PZC, as indicated in Figure \ref{fig_g-C_{3}N_{4}_mu}. This asymmetry arises from the differing shapes of the DOS observed below and above the Fermi level. Similar asymmetries have been discussed by Ochoa-Calle et al. for nitrogen-doped graphene \cite{graphene_Ndopedgraphene_Cdiff_80}. As noted by Binninger \cite{Binninger_81, Binninger_82}, the DOS can be directly related to the capacitance of the electrode-electrolyte surface \cite{Binninger_59} thus, the approximation of constant $C_{diff}$ is valid only for models with large and constant DOS. However, the correlation between the DOS and the $C_{diff}$ is proven for 2D materials and unclarified for bulk structures with low DOS near to the Fermi-level \cite{Binninger_81, Binninger_82}. Doping $g-C_{3}N_{4}$ with a single copper atom ($g-C_{3}N_{4}$-$Cu$) catalyzes the $CO_2$ reduction through its d-orbitals \cite{GCN_Cu_29, GCN_Cu_30}. The doping shifts the Fermi level toward the conduction band, resulting in electrically conductive $g-C_{3}N_{4}$-$Cu$ due to the excitable electronic states. According to Figure \ref{fig_g-C_{3}N_{4}Cu_mu_Cdiff}, it retains the non-uniform DOS of $g-C_{3}N_{4}$ and exhibits non-linear electrochemical potential and non-constant differential capacitance in relation to charge and electrode potential, respectively.
\begin{figure} [H]
      \centering
	\begin{subfigure}{0.48\linewidth}
		\includegraphics[width=\linewidth]{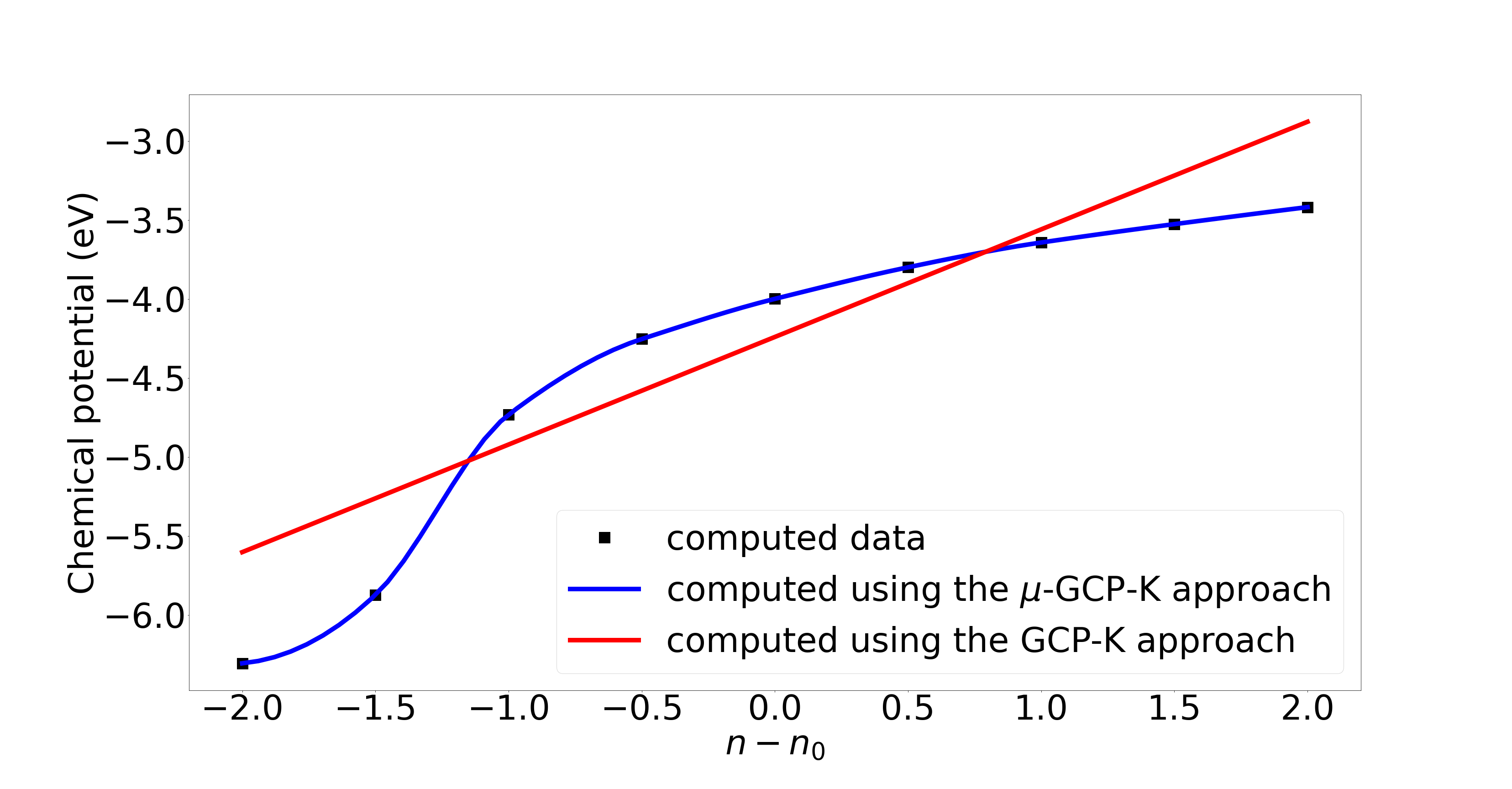}
		\caption{The $\mu \left (\Delta n\right)$ characteristic of $g-C_{3}N_{4}$-$Cu$.}
		\label{fig_g-C_{3}N_{4}Cu_mu}
	   \end{subfigure}
	   \begin{subfigure}{0.48\linewidth}
		\includegraphics[width=\linewidth]{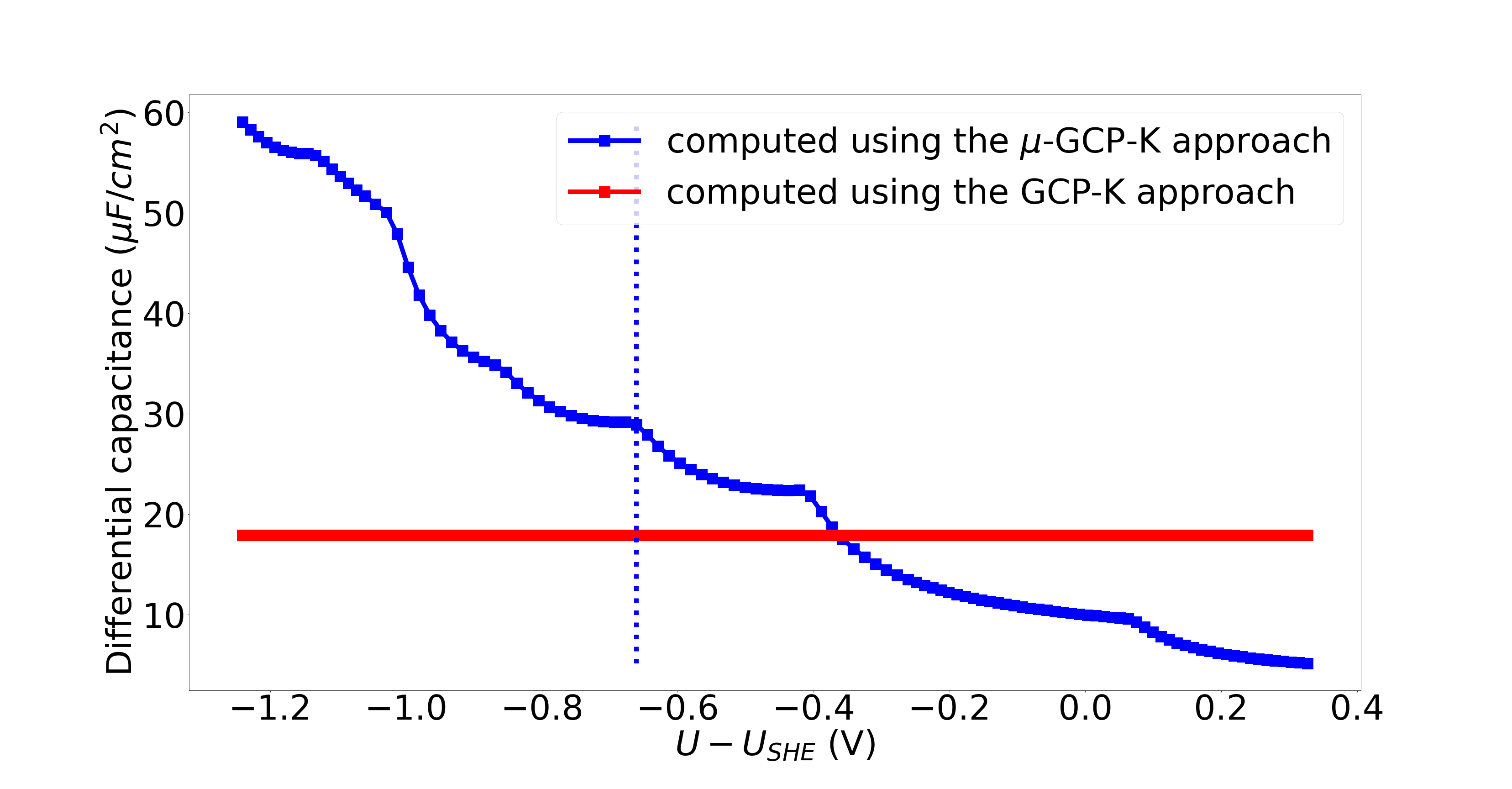}
		\caption{The $-C_{diff} \left(U-U_{SHE} \right)$ characteristic of $g-C_{3}N_{4}$-$Cu$.}
		\label{fig_g-C_{3}N_{4}Cu_Cdiff}
	    \end{subfigure}
	\caption{The electrochemical potential and differential capacitance of $g-C_{3}N_{4}$-$Cu$. Blue, vertical dashed line shows the PZC of the model in the $-C_{diff} \left(U-U_{SHE} \right)$ characteristic, calculated by the $\mu$-GCP-K approach.}
	\label{fig_g-C_{3}N_{4}Cu_mu_Cdiff}
\end{figure}
It is worth noting that differential quantities such as $C_{diff}$ are highly sensitive to variations in electronic structure, whereas thermodynamic descriptors (e.g., free energy) are integrated terms, as shown in Equation (\ref{delta_n_*}), and are expected to exhibit these changes to a lesser extent. Given that the thermodynamic properties depend on the the shape and amplitude of the potential dependent differential capacitance of $g-C_{3}N_{4}$-$Cu$ (c.f. Equation (\ref{G_U})), we are investigating  the significance of this effect on the reaction free energy of the intermediates along the possible $CO_2$ electroreduction reaction pathway leading to carbon-monoxide ($CO$)\cite{CO2RRCO_38, CO2RRCO_39, CO2RRCO_40}. Here, our goal is to investigate the effect of the potential-dependent differential capacitance on the free energies of the intermediates and products of the catalytic reaction. Assuming a proton-coupled electron transfer (PCET) \cite{PCET_41, PCET_42}, the reaction free energy, $\Delta G_{r}(*A)$ of the intermediate, $A$ can be computed as:
\begin{equation}
    \Delta G_{r}(*A) = G(*A)-G(*)-G(CO_{2})-Z \cdot \frac{1}{2} \cdot E(H_{2}).
    \label{G_r}
\end{equation}
Here, $G(*A)$ denotes the grand canonical potential of the catalyst-supported intermediate, $G(*)$ corresponds to that of the bare catalyst, while $G(CO_{2})$ to the dissolved $CO_{2}$ molecule. In the final term of Equation (\ref{G_r}), $Z$ represents the number of  $H_{2}$ atoms desorbed onto the catalyst in the reaction step of the reduction of $CO_{2}$. As we have previously discussed, accurately describing $C_{diff}(U)$ is essential for the proper computation of $G(U)$, particularly for supported structures with spatially extended electronic bands, where the amplitude of $C_{diff}$ is at least an order of magnitude larger than that of molecules with spatially localized electronic states. In the case of molecules, a more complex question arises: how to interpret their grand canonical potential under finite electrode potential and in electrochemical cells, which complicate the definition of their charge and potential state. This issue can lead to significant discrepancies in the computed stabilities of the intermediates. One way of address is to neglect the dependence of the molecule's grand canonical potential on the electrode potential assuming that when the molecules are sufficiently distant from the electrode, their grand canonical potential is independent of $U$. However, through the desorption of $CO$ from the support, we found, that in terms of the electrode - molecule distance there is a consistent discrepancy between the structure's $G(U-U_{SHE}=-0.9V)$ evaluated by assuming general and quadratic charge-free energy relation. The quadratic approximation can not precisely determine the $c$ value assumed to be equal to $F_{0}$ according to Equation (\ref{coeffs}), leading to a $\sim 0.19$ eV difference between the result of the two approaches even in asymptotic case. Next, we analyse the reaction intermediates applying the GCP-K and $\mu$-GCPK frameworks on each component, while also applying the electrode potential dependence of $G$ only to the electrode-adsorbate systems (Figure \ref{fig_Gr_g-C_{3}N_{4}Cu}).
\begin{figure} [H]
     \centering
     \begin{subfigure}[b]{0.48\textwidth}
         \centering
         \includegraphics[width=\linewidth]{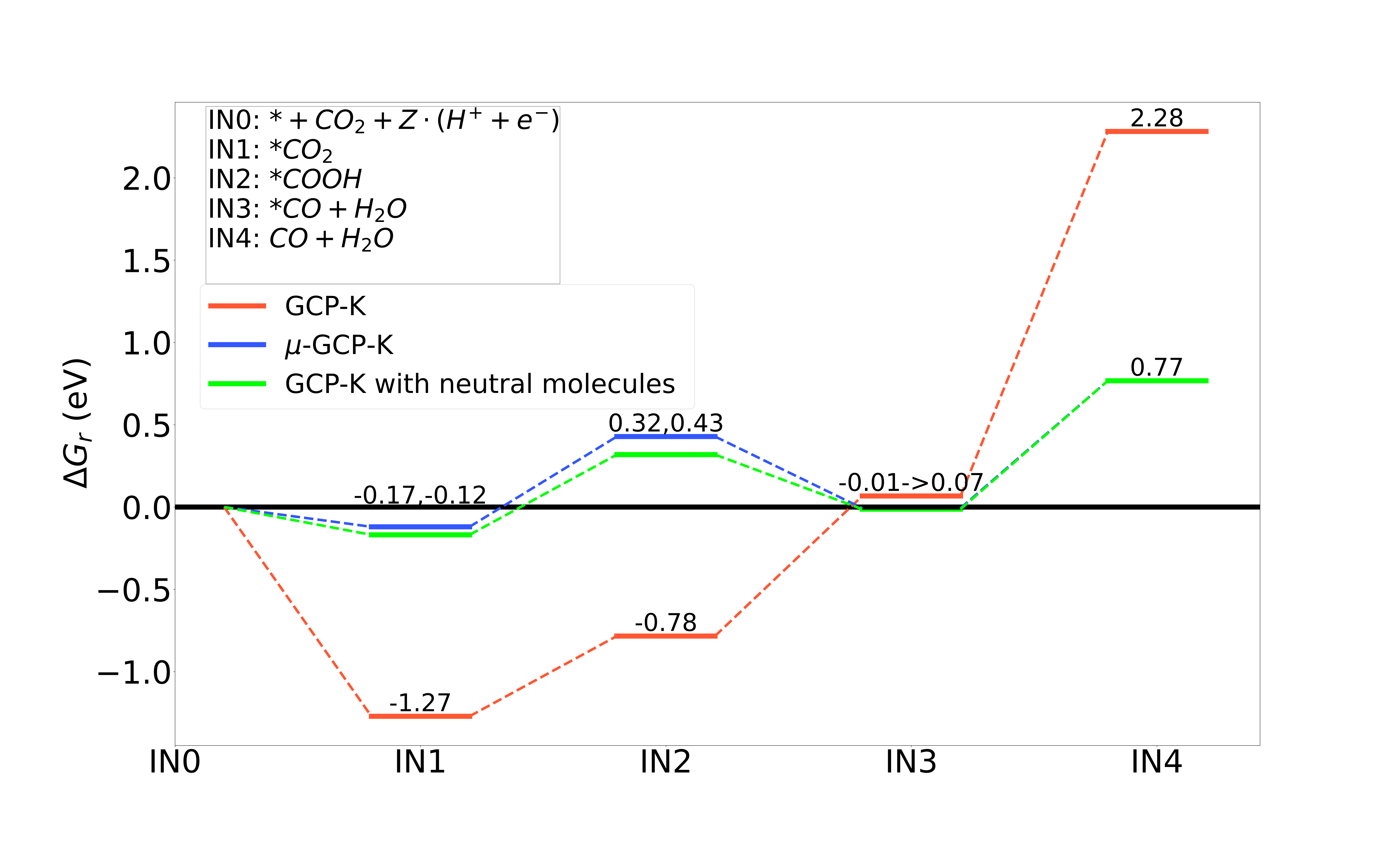}
         \caption{$U-U_{SHE}=0.0$ V.}
         \label{fig_Gr_g-C_{3}N_{4}Cu_2}
     \end{subfigure}
     \begin{subfigure}[b]{0.48\textwidth}
         \centering
         \includegraphics[width=\linewidth]{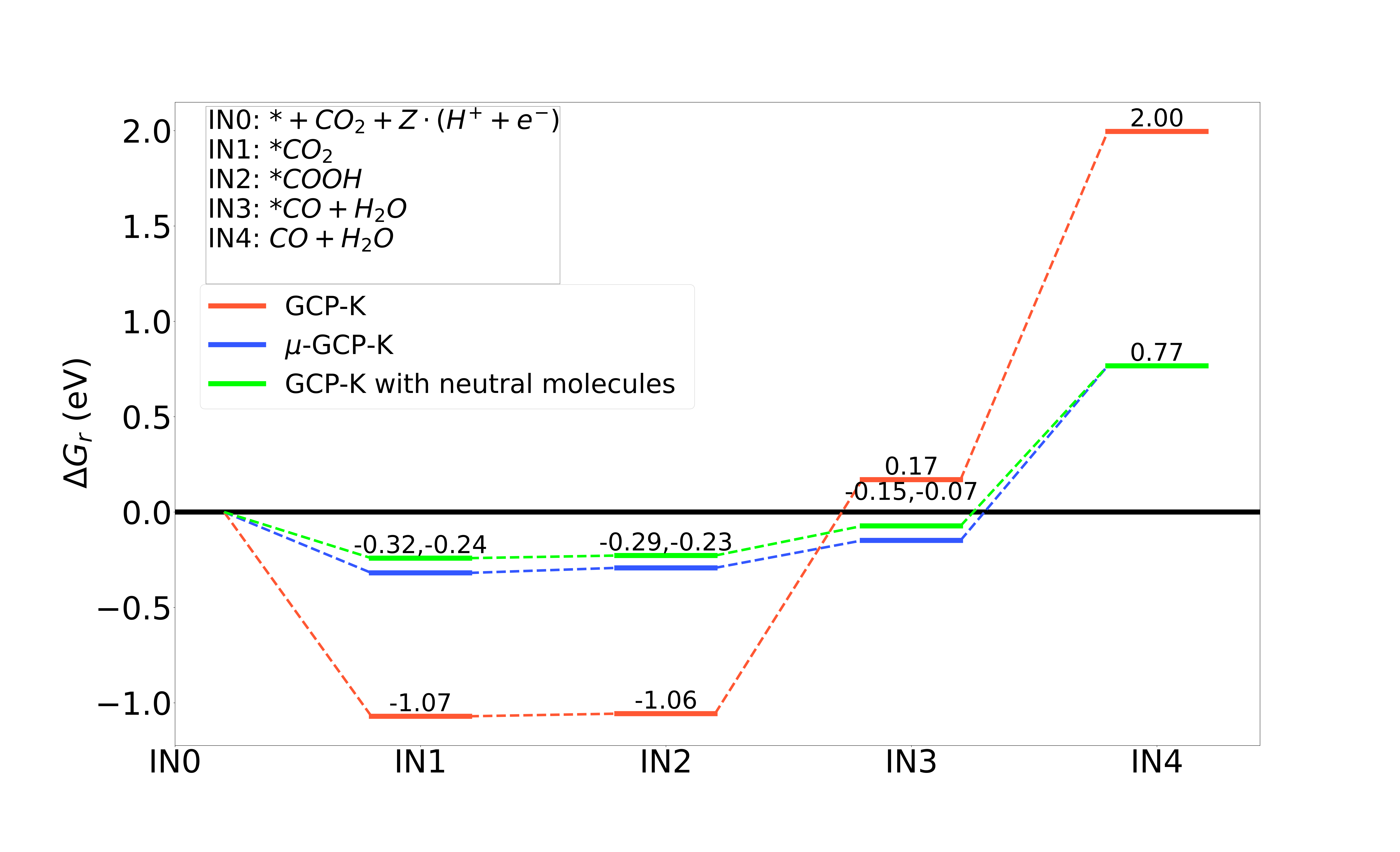}
         \caption{$U-U_{SHE}=-0.9$ V.}
         \label{fig_Gr_g-C_{3}N_{4}Cu_3}
     \end{subfigure}
        \caption{The reaction free energies of the  $CO_{2}$ reduction intermediates toward $CO$, computed at different electrode potentials (red:  quadratic $F(n)$ relation , blue: $\mu$-GCP-K), green: quadratic $F(n)$ relation applied only on electrode-adsorbate structures}
        \label{fig_Gr_g-C_{3}N_{4}Cu}       
\end{figure}
Large discrepancies are already evident in Figures \ref{fig_Gr_g-C_{3}N_{4}Cu_2} and \ref{fig_Gr_g-C_{3}N_{4}Cu_3}, where a potential of $U - U_{SHE} = 0.0$ V indicates an electron-deficient state for each intermediate, while $U - U_{SHE} = -0.9$ V  shifts to an electron-excess state. We consider the results derived from the GCP-K approach applied only to supported models to emphasize the difference compared to the $\mu$-GCP-K method, related closely to the proper computation of the differential capacitance. Quadratic potential-dependence of $G$ proposed only for supported models and the $\mu$-GCP-K result values showing similar trends. However, difference of $\sim 0.1$ eV has been observed for $*COOH$ intermediate at $U-U_{SHE}=0.0$ V. Since, $U - U_{SHE} = 0.0$ V means electron deficiency, this case has rather significance in the oxidation of $CO_{2}$. In contrast to the other intermediates, the $C_{diff}$ of $g-C_{3}N_{4}$-$Cu$-$COOH$, exhibits a greater degree of variation compared to that of $g-C_{3}N_{4}$-$Cu$ as the electrode potential is increased in the positive direction. This shows the largest discrepancy in $\Delta G_{r}$ among the intermediates at $U-U_{SHE}=0$ V as well as the major difference between the results obtained from the two methods. This highlights the importance of accurately simulating the evolution of the differential capacitance in relation to the electrode potential.\\
In this discussion, we showed the importance of accurately simulating the potential-dependent differential capacitance of electrocatalysts. We introduced the $\mu$-GCP-K method, an ab initio computational approach designed to model the effects of electrode potential on differential capacitance and other electrochemical properties. Building on GCP-K theory, the $\mu$-GCP-K method is more versatile, incorporating fewer assumptions and achieving greater accuracy without increasing the computational cost, making it particularly suitable for unconventional conductive materials. Our investigation into the catalytic properties of graphene demonstrated the necessity of basing the analysis on electronic structure to adequately capture the quantum effects that also influence differential capacitance. We also evaluated how computational accuracy impacts differential capacitance and, consequently, the stability of intermediates in the electrochemical reduction of $CO_{2}$ to $CO$ on the $g-C_{3}N_{4}$-$Cu$ catalyst. Our findings indicate that accurately computing the grand canonical potential of isolated molecules significantly affects the reaction free energy profile. Additionally, we demonstrated the possible consequences of the non-constant differential capacity on the stability of key intermediates in electrocatalysis. Overall, the $\mu$-GCP-K method is well-suited for precisely simulating both traditional metallic electrode models and those exhibiting non-metallic electronic structure effects. We believe that this universality is crucial for the design and modification of novel electrocatalysts.

\section*{Acknowledgement}
    M.G. is grateful to the Cooperative Doctoral Programme for Doctoral Scholarships of the Hungarian National Research, Development and Innovation Office (KDP-23, grant number: C2273140). We thank the Hungarian Government and the European Union, Grant/Award Number: VEKOP‐2.1.1‐15‐2016‐00114 for supporting the GPU computational resources. This project was supported the European Union’s Horizont 2020 research and innovation programme under the Marie Sklodowska-Curie grant agreement no. 955650. During the preparation of this manuscript, the authors used ChatGPT to improve the grammar and coherence of the text. All material was subsequently reviewed and revised by the authors, who take full responsibility for the content of the publication.

\begin{suppinfo}

The Supporting Information is available free of charge at...
\begin{itemize}
  \item Computational details:
  \begin{itemize}
    \item Computational parameters - Section 1.1 
    \item Parameter tests on vacuum size, k-point set and spin-polarization - Section 1.2 
    \end{itemize}  
  \item The application of GCP-K approaches:
  \begin{itemize}
    \item Computed GCP-K parameters on graphene - Section 2.1
    \item Computed GCP-K parameters, free energies, chemical potentials and differential capacitances on g-$C_{3}N_{4}-Cu$ and the intermediates of the $CO$ reaction pathway of the $CO_{2}$ reduction
    \end{itemize}  
  \item Details on studying the intermediates of the g-$C_{3}N_{4}-Cu$ catalysed $CO_{2}$ electrochemical reduction to $CO$:
    \begin{itemize}
    \item Grand canonical potential of desorbed $CO$ from g-$C_{3}N_{4}-Cu$
    \item Grand canonical potentials, enthalpy and entropy contributions of reaction intermediates, reaction free energies, differences of differential capacitances of reaction intermediates.
    \end{itemize} 
\end{itemize}

\end{suppinfo}

\section*{Data Availability}
All data not presented in the Supporting Information have been uploaded to the ZENODO repository at DOIs: https://doi.org/10.5281/zenodo.16964422

\bibliography{references}

\end{document}